\input{aipcheck}

\documentclass[
    ,final            
  ]
  {aipproc}
\layoutstyle{8x11single}

\usepackage{amssymb}

\begin{document}

\title{Investigation of the radio wavefront of air showers with LOPES measurements and CoREAS simulations}


\author{F.G.~Schr\"oder\footnote{Corresponding author \newline Email address: frank.schroeder@kit.edu} }{
  address={Karlsruhe Institute of Technology (KIT), Germany}
}

\author{W.D.~Apel}{
address={Karlsruhe Institute of Technology (KIT), Germany}
}

\author{J.C.~Arteaga-Velazquez}{
address={Universidad Michoacana, Morelia, Mexico}
}

\author{L.~B\"ahren}{
address={Radboud University Nijmegen, Department of Astrophysics, The Netherlands}
}

\author{K.~Bekk}{
address={Karlsruhe Institute of Technology (KIT), Germany}
}

\author{M.~Bertaina}{
address={Dipartimento di Fisica dell' Universit\`a Torino, Italy}
}

\author{P.L.~Biermann}{
address={Max-Planck-Institut f\"ur Radioastronomie Bonn, Germany}
}

\author{J.~Bl\"umer}{
address={Karlsruhe Institute of Technology (KIT), Germany}
}

\author{H.~Bozdog}{
address={Karlsruhe Institute of Technology (KIT), Germany}
}

\author{I.M.~Brancus}{
address={National Institute of Physics and Nuclear Engineering, Bucharest, Romania}
}

\author{E.~Cantoni}{
address={Dipartimento di Fisica dell' Universit\`a Torino, Italy}
}

\author{A.~Chiavassa}{
address={Dipartimento di Fisica dell' Universit\`a Torino, Italy}
}

\author{K.~Daumiller}{
address={Karlsruhe Institute of Technology (KIT), Germany}
}

\author{V.~de~Souza}{
address={Universidad S\~ao Paulo, Inst. de F\'{\i}sica de S\~ao Carlos, Brasil}
}

\author{F.~Di~Pierro}{
address={Dipartimento di Fisica dell' Universit\`a Torino, Italy}
}

\author{P.~Doll}{
address={Karlsruhe Institute of Technology (KIT), Germany}
}

\author{R.~Engel}{
address={Karlsruhe Institute of Technology (KIT), Germany}
}

\author{H.~Falcke}{
address={Radboud University Nijmegen, Department of Astrophysics, The Netherlands}
,altaddress={ASTRON, Dwingeloo, The Netherlands}
}

\author{B.~Fuchs}{
address={Karlsruhe Institute of Technology (KIT), Germany}
}

\author{D.~Fuhrmann}{
address={Universit\"at Wuppertal, Fachbereich Physik, Germany}
}
\author{H.~Gemmeke}{
address={Karlsruhe Institute of Technology (KIT), Germany}
}

\author{C.~Grupen}{
address={Universit\"at Siegen, Fachbereich Physik, Germany}
}

\author{A.~Haungs}{
address={Karlsruhe Institute of Technology (KIT), Germany}
}

\author{D.~Heck}{
address={Karlsruhe Institute of Technology (KIT), Germany}
}

\author{J.R.~H\"orandel}{
address={Radboud University Nijmegen, Department of Astrophysics, The Netherlands}
}

\author{A.~Horneffer}{
address={Max-Planck-Institut f\"ur Radioastronomie Bonn, Germany}
}

\author{D.~Huber}{
address={Karlsruhe Institute of Technology (KIT), Germany}
}

\author{T.~Huege}{
address={Karlsruhe Institute of Technology (KIT), Germany}
}

\author{P.G.~Isar}{
address={Institute for Space Sciences, Bucharest, Romania}
}

\author{K.-H.~Kampert}{
address={Universit\"at Wuppertal, Fachbereich Physik, Germany}
}

\author{D.~Kang}{
address={Karlsruhe Institute of Technology (KIT), Germany}
}

\author{O.~Kr\"omer}{
address={Karlsruhe Institute of Technology (KIT), Germany}
}

\author{J.~Kuijpers}{
address={Radboud University Nijmegen, Department of Astrophysics, The Netherlands}
}

\author{K.~Link}{
address={Karlsruhe Institute of Technology (KIT), Germany}
}

\author{P.~{\L}uczak}{
address={National Centre for Nuclear Research, Department of Astrophysics, {\L}\'{o}d\'{z}, Poland}
}

\author{M.~Ludwig}{
address={Karlsruhe Institute of Technology (KIT), Germany}
}

\author{H.J.~Mathes}{
address={Karlsruhe Institute of Technology (KIT), Germany}
}

\author{M.~Melissas}{
address={Karlsruhe Institute of Technology (KIT), Germany}
}

\author{C.~Morello}{
address={INAF Torino, Osservatorio Astrofisico di Torino, Italy}
}

\author{J.~Oehlschl\"ager}{
address={Karlsruhe Institute of Technology (KIT), Germany}
}

\author{N.~Palmieri}{
address={Karlsruhe Institute of Technology (KIT), Germany}
}

\author{T.~Pierog}{
address={Karlsruhe Institute of Technology (KIT), Germany}
}

\author{J.~Rautenberg}{
address={Universit\"at Wuppertal, Fachbereich Physik, Germany}
}

\author{H.~Rebel}{
address={Karlsruhe Institute of Technology (KIT), Germany}
}

\author{M.~Roth}{
address={Karlsruhe Institute of Technology (KIT), Germany}
}

\author{C.~R\"uhle}{
address={Karlsruhe Institute of Technology (KIT), Germany}
}

\author{A.~Saftoiu}{
address={National Institute of Physics and Nuclear Engineering, Bucharest, Romania}
}

\author{H.~Schieler}{
address={Karlsruhe Institute of Technology (KIT), Germany}
}

\author{A.~Schmidt}{
address={Karlsruhe Institute of Technology (KIT), Germany}
}

\author{S.~Schoo}{
address={Karlsruhe Institute of Technology (KIT), Germany}
}


\author{O.~Sima}{
address={University of Bucharest, Department of Physics, Romania}
}

\author{G.~Toma}{
address={National Institute of Physics and Nuclear Engineering, Bucharest, Romania}
}

\author{G.C.~Trinchero}{
address={INAF Torino, Osservatorio Astrofisico di Torino, Italy}
}

\author{A.~Weindl}{
address={Karlsruhe Institute of Technology (KIT), Germany}
}

\author{J.~Wochele}{
address={Karlsruhe Institute of Technology (KIT), Germany}
}

\author{J.~Zabierowski}{
address={National Centre for Nuclear Research, Department of Astrophysics, {\L}\'{o}d\'{z}, Poland}
}

\author{J.A.~Zensus}{
address={Max-Planck-Institut f\"ur Radioastronomie Bonn, Germany}
}

\begin{abstract}
We investigated the radio wavefront of cosmic-ray air showers with LOPES measurements and CoREAS simulations: the wavefront is of approximately hyperbolic shape and its steepness is sensitive to the shower maximum. For this study we used $316$ events with an energy above $0.1\,$EeV and zenith angles below $45^\circ$ measured by the LOPES experiment. LOPES was a digital radio interferometer consisting of up to $30$ antennas on an area of approximately $200\,$m$\times200\,$m at an altitude of $110\,$m above sea level. Triggered by KASCADE-Grande, LOPES measured the radio emission between $43\,$ and $74\,$MHz, and our analysis might strictly hold only for such conditions. Moreover, we used CoREAS simulations made for each event, which show much clearer results than the measurements suffering from high background. A detailed description of our result is available in our recent paper published in \emph{JCAP09(2014)025} \cite{2014ApelLOPES_Wavefront}. The present proceeding contains a summary and 
focuses on some additional aspects, e.g., the asymmetry of the wavefront: According to the CoREAS simulations the wavefront is slightly asymmetric, but on a much weaker level than the lateral distribution of the radio amplitude.
\end{abstract}

\maketitle

\section{Introduction}
There are several motivations bringing the wavefront of the radio signal emitted by air showers into the focus of interest: First, the wavefront is of principle interest, since the wavefront is linked to the process of the radio emission. Thus, understanding the wavefront means understanding the relevant parts of the emission process. Second, the wavefront can be used for the reconstruction of the cosmic-ray composition, since the steepness of the wavefront is correlated to the distance to the shower maximum, which itself statistically depends on the type of the primary particle. Third, the wavefront can be a technical instrument to generally improve the radio detection of air-showers, e.g., by a better reconstruction of the shower geometry, or by a better discrimination of air showers against background events.


For the analysis of the radio wavefront we used 316 LOPES events triggered by the co-located KASCADE array \cite{AntoniApelBadea2003}, which fulfill the following criteria of the KASCADE reconstruction: energy $E > 10^{17}\,$eV, zenith angle $\theta < 45^\circ$, core position within $90\,$m from the center of KASCADE. Thus, for all those selected events there are antennas in different azimuthal directions, and the asymmetry of the wavefront partially averages out. Moreover, the events had to show a clear radio signal in the cross-correlation beam \cite{FalckeNature2005}, and we excluded thunderstorm events \cite{2011ApelLOPESthunderstorm}. The analysis of the LOPES events followed our standard reconstruction pipeline, with the one exception that we assumed a hyperbolic wavefront for the beamforming instead of a spherical or conical wavefront. See references \cite{2014ApelLOPES_Wavefront, 2013ApelLOPESlateralComparison, HuegeARENA_LOPESSummary2010} for more details on the standard LOPES analysis.

This means that our results are strictly valid only for LOPES conditions, i.e., in particular for the east-west polarization component of the radio signal in the effective frequency range of $43-74\,$MHz, the LOPES altitude of $110\,$m above sea level, the geomagnetic field at Karlsruhe, Germany, and the distance range up to $200\,$m. Nevertheless, a preliminary analysis based on a CoREAS simulation indicates that the wavefront has still the same, approximately hyperbolic shape when extending the bandwidth from very low frequencies up to $100\,$MHz.

For each LOPES event we produced two CoREAS simulations \cite{HuegeARENA_Coreas_2012} (one with proton, one with an iron nucleus as primary particle) using the KASCADE reconstruction of the air shower as input parameters. In addition, we simulated a few vertical showers with an artificial grid of antennas which are aligned exactly in east, west, north and south direction, to study the asymmetry of the wavefront due to the interference of the dominant geomagnetic and the sub-dominant Askaryan effect. For the CoREAS simulations the geometry is known exactly. Thus, unlike to the LOPES measurements, we have not applied the beamforming method to reconstruct the wavefront, but instead performed a simple fit to the pulse arrival time at each antenna position (time of the maximum in the east-west component of the radio signal filtered to the effective bandwidth of LOPES).

\section{Results}
\begin{figure}[t]
\centering
\includegraphics[width=0.48\textwidth]{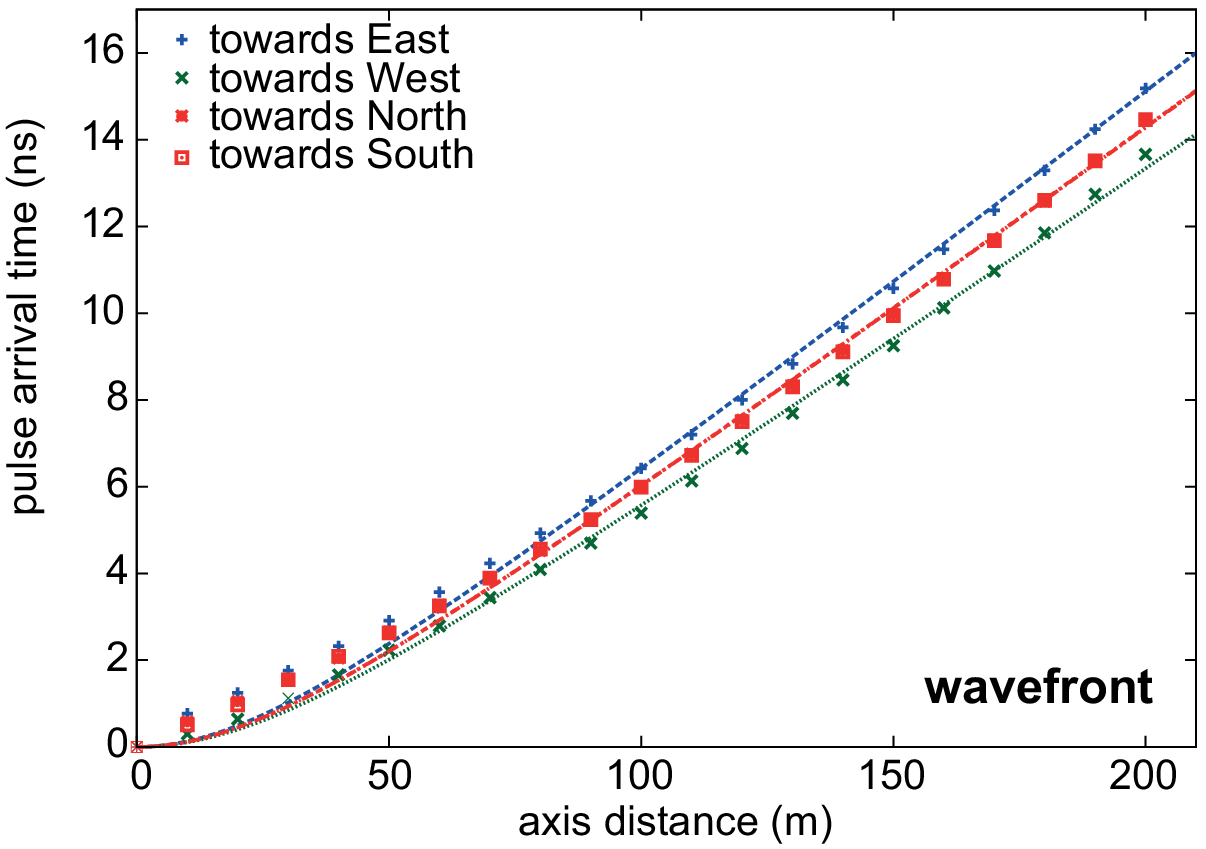}
  \hskip 0.2 cm
\includegraphics[width=0.48\textwidth]{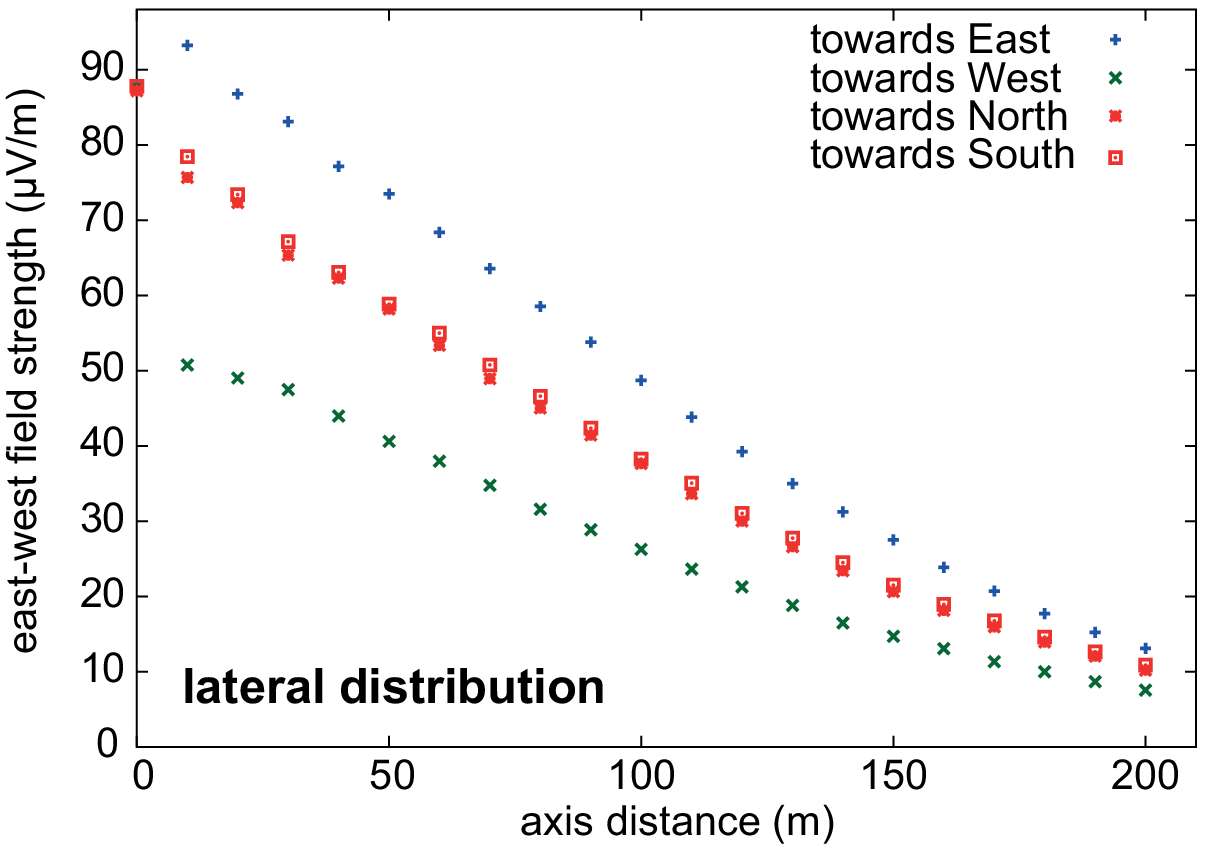}
\caption{Asymmetry of the wavefront (left) and the lateral distribution (right) of a vertical shower initiated by a $10^{17}\,$eV proton simulated with CoREAS in different azimuthal directions from the core, and hyperbolic fits to the wavefronts.} \label{fig_assymetry}
\end{figure}

The CoREAS simulations show that the radio wavefront is asymmetric in a similar manner to already known asymmetry of the lateral distribution \cite{AugerAERApolarization2014, CODALEMAMarinICRC2011}, but on a much weaker scale as shown in figure \ref{fig_assymetry}. Thus, for most purposes a symmetric wavefront is a good approximation, especially, since the asymmetry will partially average out in real measurements which have antennas in several azimuthal directions from the shower core.

Therefore, we compared only different models for symmetric wavefronts with each other. In the simulations, the hyperbolic wavefront is clearly favored versus the spherical and the conical wavefront, although a conical wavefront might be a sufficient approximation for larger distances to the shower axis ($d \gtrsim 50\,$m). The measurements are compatible with this result, but the situation is not so clear, since there is significant background. Nevertheless, the hyperbolic wavefront is slightly favored by the measurements, too. For example, the average deviation between the arrival directions reconstructed by KASCADE and by LOPES is smallest when using the hyperbolic assumption for the beamforming. It is $(0.683 \pm 0.026)^\circ$ when assuming a spherical wavefront, $(0.637 \pm 0.022)^\circ$ for a conical wavefront, and  $(0.622 \pm 0.023)^\circ$ for the hyperbolic wavefront.

The symmetric, hyperbolic wavefront is described by the following formula for the arrival time $\tau$ relative to the arrival time at the shower core:
\begin{equation}
c \, \tau(d,z_s) = \\ \sqrt{(d\sin\rho)^2+(c\cdot b)^2} + z_s\cos\rho + c\cdot b
\label{eq_hyperbola}
\end{equation}
where $d$ is the orthogonal distance to the shower axis, $z_s$ is the orthogonal distance to the plane (with $z_s = 0$ at the shower core), $c$ is the speed of light, $\rho$ is the angle between the asymptotic cone of the hyperboloid and the shower plane, and $b$ is the offset between the hyperboloid an the apex of the asymptotic cone. Thus, the hyperbolic wavefront has two parameters, namely $b$ and $\rho$. Both parameters are correlated to some degree, and we fixed $b$ to $-3\,$ns for the reconstruction of the wavefront. By fixing $b$, we improve the correlation of the remaining free parameter $\rho$ with the distance to the shower maximum: The larger the distance to the shower maximum, the flatter the wavefront and the smaller $\rho$.

The average values reconstructed for $\rho$ are: $(0.0210 \pm 0.0055)\,$rad for the LOPES measurements,  $(0.0216 \pm 0.0036)\,$rad for the proton simulations, and $(0.0194 \pm 0.0030)\,$rad for the   
iron simulations, i.e., the wavefront deviates from the shower plane by more than $1^\circ$.

The distance to the shower maximum depends on its atmospheric depth $X_\mathrm{max}$ and the zenith angle $\theta$ of the showers. To reconstruct $X_\mathrm{max}$ from the cone angle $\rho$, the zenith angle dependence of $\rho$ has to be corrected. This correction is difficult to calculate, because in first order the shower is not a point source, but an extended line source (which only is an approximation, too, because of the lateral extension of the shower). Thus, we have determined the zenith dependence of $\rho$ empirically, and correct for it with a simple parametrization. After the correction, the cone angle $\rho$ is approximately linearly correlated with $X_\mathrm{max}$:

\begin{equation}
X_\mathrm{max} \approx 25,000\,\mathrm{g/cm}^2 \cdot \rho/\mathrm{rad} \cdot \cos^{-1.5} \theta
\label{eq_XmaxReconstruction}
\end{equation}
The coefficients have been determined from the CoREAS simulations: The coefficients for proton and iron simulations are similar within a few percent, but not identical. The zenith dependence is the same for LOPES and CoREAS (within uncertainties). However, the proportionality coefficient cannot be checked experimentally, since LOPES/KASCADE do not feature any reference measurement of $X_\mathrm{max}$.

\begin{figure}[t]
  \centering
  \includegraphics[width=0.49\textwidth]{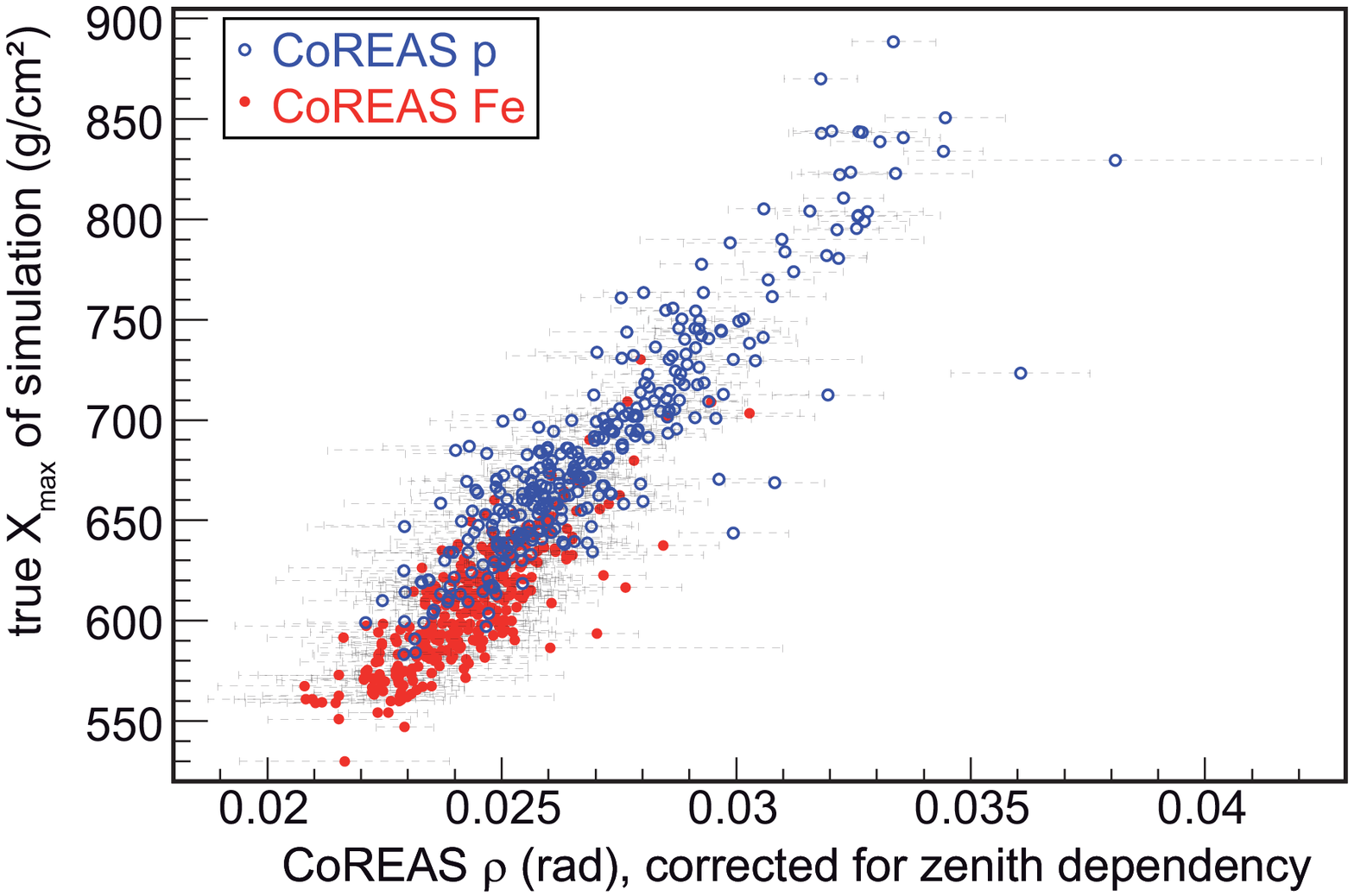}
   \hskip 0.2 cm
  \includegraphics[width=0.47\textwidth]{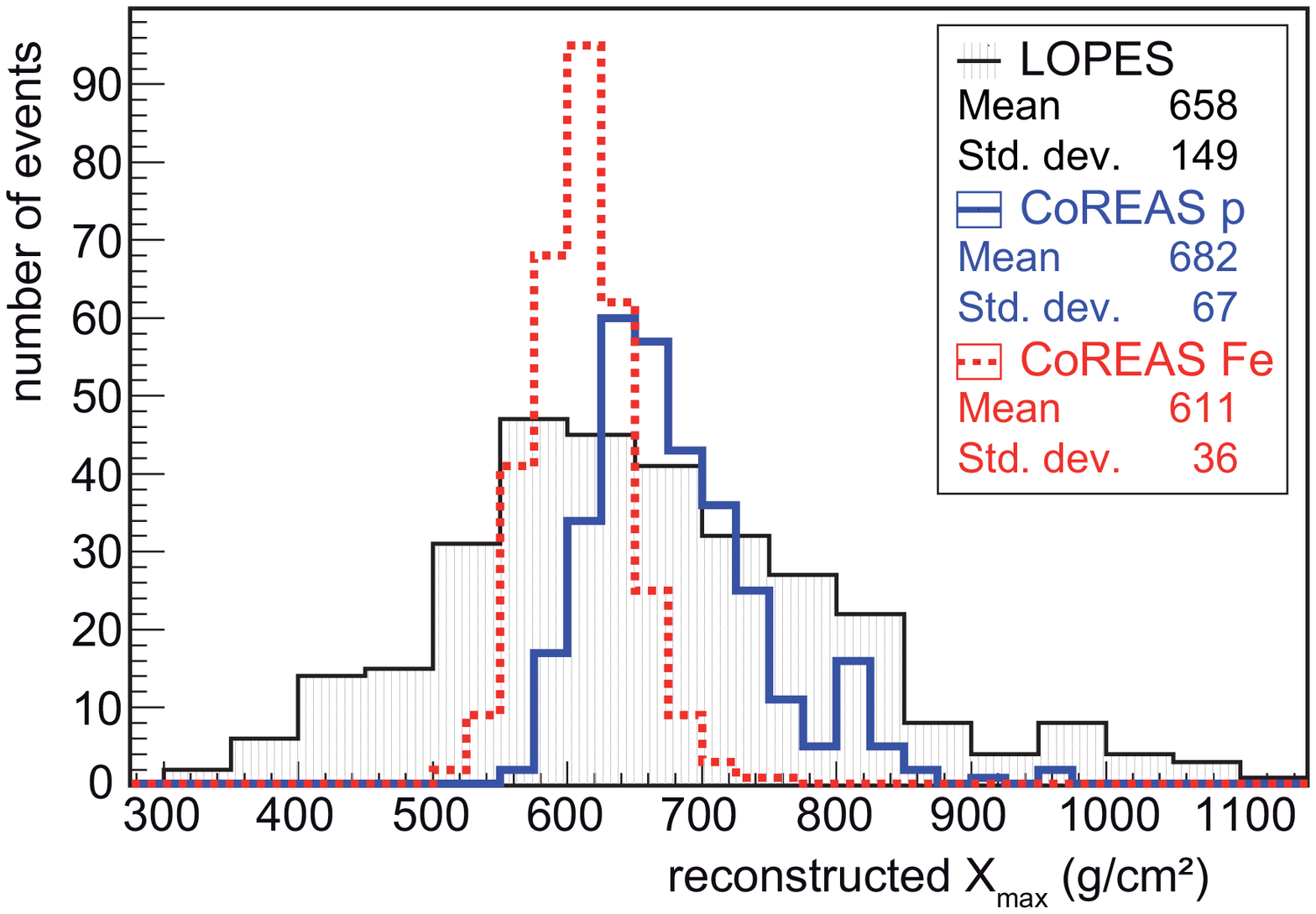}
  \caption{Left: correlation of the cone angle $\rho$ after correction for the shower inclination with the true atmospheric depth of the shower maximum $X_\mathrm{max}$. Right: $X_\mathrm{max}$ reconstructed by making use of the linear correlation in the left plot (published in \cite{2014ApelLOPES_Wavefront}).}
   \label{fig_Xmax}
\end{figure}

Figure \ref{fig_Xmax} shows the comparison of the reconstructed and the true $X_\mathrm{max}$ for the simulations. The reconstruction is precise to approximately $25\,$g/cm\textsuperscript{2}. This means that under ideal conditions (no background, known shower geometry), light and heavy nuclei can be well separated. When applying the reconstruction derived from the simulations to the measurements, the result seems to be reasonable, but the precision of the measurements is significantly worse: most of the width in the distribution of the LOPES $X_\mathrm{max}$ values is due to measurement uncertainties. Since LOPES features a ns-precise time calibration \cite{SchroederTimeCalibration2010}, we conclude that the main problem is the external radio background at the experimental site close to Karlsruhe, which causes significant measurement uncertainties for the pulse time \cite{SchroederNoise2010}.

The remaining uncertainty of $25\,$g/cm\textsuperscript{2} in the simulations causing the spread in figure \ref{fig_Xmax} (left) is not due to the fixed the offset parameter $b$. But it might be due to the asymmetry of the wavefront, or because even a hyperboloid is not the exact shape of the wavefront, but only an approximation. Consequently, a more detailed investigation of the wavefront might be necessary when aiming at a $X_\mathrm{max}$ resolution of better than $25\,$g/cm\textsuperscript{2}.

\section{Conclusion}
As seen clearly for the CoREAS simulations, as statistically indicated for the LOPES measurements, and in agreement with recent LOFAR results \cite{CorstanjeLOFAR_wavefront2014}, the radio wavefront of air showers can be better described by a hyperboloid than by a plane, a sphere, or a cone. Although the wavefront is slightly asymmetric, according to the simulations, a symmetric wavefront is a sufficient approximation, at least for the currently achieved measurement precision and for distances $\lesssim 200\,$m to the shower axis. 

The steepness of the wavefront, namely the cone angle $\rho$ can be used to reconstruct $X_\mathrm{max}$ and to statistically study the cosmic-ray composition. The accuracy likely can be improved by combining the wavefront method with other independent measurements of the same events. For example, the slope of the lateral distribution of the radio amplitude is also sensitive to $X_\mathrm{max}$ \cite{2012ApelLOPES_MTD, 2014ApelLOPES_MassComposition}.

Finally, the improved knowledge of the wavefront leads to a better reconstruction of the shower geometry. Provided sufficient measurement precision for the arrival times, wavefront measurements could be used to distinguish air showers from other radio sources, e.g., air planes: because many background sources are in good approximation point sources, background events usually should have a spherical wavefront and not a hyperbolic one.

\begin{theacknowledgments}
LOPES and KASCADE-Grande have been supported by the German Federal Ministry of Education and Research. KASCADE-Grande is partly supported by the MIUR and INAF of Italy, the Polish Ministry of Science and Higher Education and by the Romanian Authority for Scientific Research UEFISCDI (PNII-IDEI grant 271/2011). This research has been supported by grant number VH-NG-413 of the Helmholtz Association. The present study is supported by the `Helmholtz Alliance for Astroparticle Physics - HAP` funded by the Initiative and Networking Fund of the Helmholtz Association, Germany.

\end{theacknowledgments}

\bibliographystyle{aipproc}
\bibliography{arena2014}

\end{document}